\documentclass{hermann}
\usepackage{lipsum}
\usepackage{hyperref}
\usepackage[english]{babel}
\usepackage[
backend=bibtex,
style=authoryear,
citestyle=authoryear-comp,
eprint=false,
isbn=false,
url=false, 
doi=false, 
natbib=true,
language=english,
hyperref=auto
]{biblatex}
\renewbibmacro{in:}{%
  \ifentrytype{article}{}{\printtext{\bibstring{in}\intitlepunct}}}

\bibliography{Henon}
\AtEveryBibitem{
\clearfield{note}
\clearfield{day}%
\clearfield{month}%
\clearfield{endday}%
\clearfield{endmonth}%
}    
\usepackage{url}

\newcommand\crm{\cr\noalign{\medskip}}

\newcommand\Frac[2]{{{\displaystyle\strut#1}\over{\displaystyle\strut#2}}}

\newcommand\be{\begin{equation}}
\newcommand\ee{\end{equation}}
\def\m@th{\mathsurround=0pt}
\newcommand\EQM[1]{\vcenter{\normalbaselines\m@th
    \ialign{${\displaystyle ##}$\hfil&&\ ${\displaystyle ##}$\hfil\crcr
    \mathstrut\crcr\noalign{\kern-\baselineskip}
    \noalign{\smallskip}
    #1\crcr\mathstrut\crcr\noalign{\kern-\baselineskip}}}}

\title{Michel Hénon and the Stability of the Solar System}
\author{Jacques Laskar}

\begin{document}

\maketitle{}

\tableofcontents{}

\chapter*{Michel Hénon and the Stability of the Solar System}

\section{Introduction}
My first meeting with Michel Hénon was at the occasion of  a conference organised 
by Claude Froeschlé on  dynamical interactions in the Solar System at Aussois in the french Alps. 
During this conference, one free afternoon was devoted to a hike in the mountain. One should 
never  stress enough the importance of these free times in scientific conferences, 
when there are no  talks, and  thus when people  can discuss freely about their projects. 
At  the time, I was thinking of  starting a  study group for the reading of the "Méthodes Nouvelles 
de la Mécanique Céleste", the masterpiece of  Henri  Poincaré. As stated by  Ivar Ekeland \citep{Ekel1987a},
it appeared that this fundamental work was often quoted but never read. 
Discussing of this project to colleagues, I had often  some reaction saying that, it was 
interesting, but too difficult. On our way back from the shelter of the  Fond d'Aussois, in 
this nice autumn afternoon, I had a long discussion with Michel Hénon who strongly encouraged me
to continue in this project. Later on, I had an enthusiastic support from Alain Chenciner, 
and we started this three year project on the reading of Poincaré. This study, which could appear as 
unusual at present for a young researcher, who is normally pushed to publish as much as possible, 
instead of spending time reading a  100 years old book, had profound implications as 
it was at the origin of the creation of our multidisciplinary research group (Astronomie et Systèmes Dynamiques) 
in the Bureau des Longitudes, gathering astronomers and mathematicians. Moreover,  the numerous discussions that we had 
during these lectures of Poincaré were of fundamental help for me for the understanding 
of the chaotic behaviour of the Solar System \citep{Lask1989a,Lask1990a}.

\section{Historical background}
The question of the stability of the Solar System is a very old question that was already raised by 
 Newton in his Opticks volume \citep{Newt1706a,Newt1718a}. He envisioned the presence of God in the 
 apparent ordering of the planets, but on the opposite, he admitted that the perturbations among the planets 
 may lead to large  instabilities   in a way that a divine intervention would become 
 necessary. This was  highly contested by the rival of Newton, Leibniz, who  found that 
 suggesting that the clockwork of God  does not achieve indefinite stability and needs
 to be mended from time to time, is contesting the  power of God. The question of the stability of 
 the Solar System was thus a question of limiting or not the divine power, and was one of the 
 main scientific questions of the XVIIIth century. It was even more important 
 as after Kepler already in 1625, Halley had found that it was necessary to add some empirical linear drift in 
 the mean motions of Jupiter and Saturn in order to retrieve the Chaldean observations transmitted by 
 Ptolemea  \citep[see][]{Lask2013a}. Jupiter was going towards the Sun while Saturn was escaping from the Sun. 
 It was thus natural for Newton to  question the stability of the Solar System, and it was not until the end 
 of the XVIII century that the question found a satisfying answer. Indeed, Laplace demonstrated that 
 at first order, the mean values of the semi-major axis are constant. He also showed that 
 the apparent variations of the 
 mean motion over the past two millennium were due to an oscillation of the longitude resulting from the 
 the proximity of the 5:2 resonance between the motions of Jupiter and Saturn. 
 Laplace thus demonstrated that  no  instabilities could arise from the varying size of the planetary orbits. 
 Lagrange and Laplace completed the first "proof of the stability of the Solar System" by demonstrating 
 that at first order in the masses, eccentricities and inclinations, the planetary orbits 
 suffer precession motion of their perihelion and node, and  oscillations of their eccentricity and inclination, 
 but of small values that do not allow for collision (see \citep{Lask2013a} for a detailed account). 
 
 As stated by \citep{Poin1897a}, many other "proofs" of the stability of the Solar System will 
 follow. It does not mean that the previous ones were insufficient, but that they 
 addressed only to some approximation of the problem, which may be far from   reality. 
 Poincaré himself demonstrated that the three body problem is not integrable, and that 
 the series used by the astronomers were divergent, thus not allowing 
 to decide for the stability of the Solar System in infinite time. He also exhibited  the very 
 complex behaviour that can have the solutions of the three body problem, 
 describing in great details  the intricate imbrication of the stable and unstable varieties in 
 the vicinity of an hyperbolic fixed point. It is in these regions that originate the chaotic 
 solutions that are very sensitive to  small changes of their initial conditions \citep{Poin1899a}.
 In fact, Poincaré did not seem to think that his results apply to the physical Solar System, 
 and in   \citep{Poin1897a}, he assumed that the small tidal dissipative effects were 
 more important than the chaotic diffusion, and thus could stabilise the system 
 which would end  in a state of rigid rotation, where all the planets would be synchronised,
 and  synchronised with the rotation of the Sun. Although the  tidal effects do exist, 
 it can be estimated that their effect is much smaller than what Poincaré forecasted, 
 and we can for example   estimate to the order of  $10^{15}$ billion  years, 
 the time needed to circularise the orbit of Jupiter \citep{Corr2011a}.

 In addition, although Poincaré demonstrated that there cannot be a domain of initial 
 conditions with integrable, regular solutions,  he did not excluded that there could be some 
 special conditions under which the perturbation series would converge, and thus the 
 solutions would be regular   \citep{Poin1892a}. Sixty years later, \citet{Kolm1954a} actually 
 demonstrated the possibility of regular quasiperiodic solutions in a perturbed Hamiltonian system.
 The convergence of the series is obtained under the conditions that the perturbation is small and 
  that the frequencies of the 
 system verify some Diophantine condition, which for two degrees of freedom systems 
 means that the frequency ratio is far from the rationals \citep[see][]{Chie2006a,Duma2014a}. 
 
 \section{Regularity and chaos in Hamiltonian systems}
 A first application of Kolmogorov theorem  to  a planetary problem was  later on provided by 
 \citep{Arno1963a,Arno1963b}, in a planar case, for a ratio of the semi-major axis 
 close to zero. This impressive result was considered as another  "demonstration" of the stability 
 of the Solar System, and is quoted as such by Jurgen Moser in a general audience paper \citep{Mose1978a}.
Moser himself contributed to the theory with an alternate demonstration of Kolmogorov theorem, valid 
for a larger class of functions \citep{Mose1962a}. These results are since known as KAM theory (Kolmogorov-Arnold-Moser).
One of the main result of the KAM theory, is to state that regular quasiperiodic solutions may exist 
in places where it was previously thought that all solutions will be unstable. 
Moreover,  as the perturbation goes to zero, the density of the regular solutions goes to unity. 
This was at first considered as a mathematical curiosity, but very soon, Michel Hénon 
was the first to show numerical evidence of this behaviour of conservative (Hamiltonian) systems,
in a celebrated paper on the dynamics of stars in a galactic potential \citep{HenoHeil1964a}. 
It appeared clearly that the general setting described by the KAM theory, with an intricate 
mixture of regular and chaotic trajectories was actually present in realistic physical  models.

\begin{figure}[h]
\includegraphics[width=7cm]{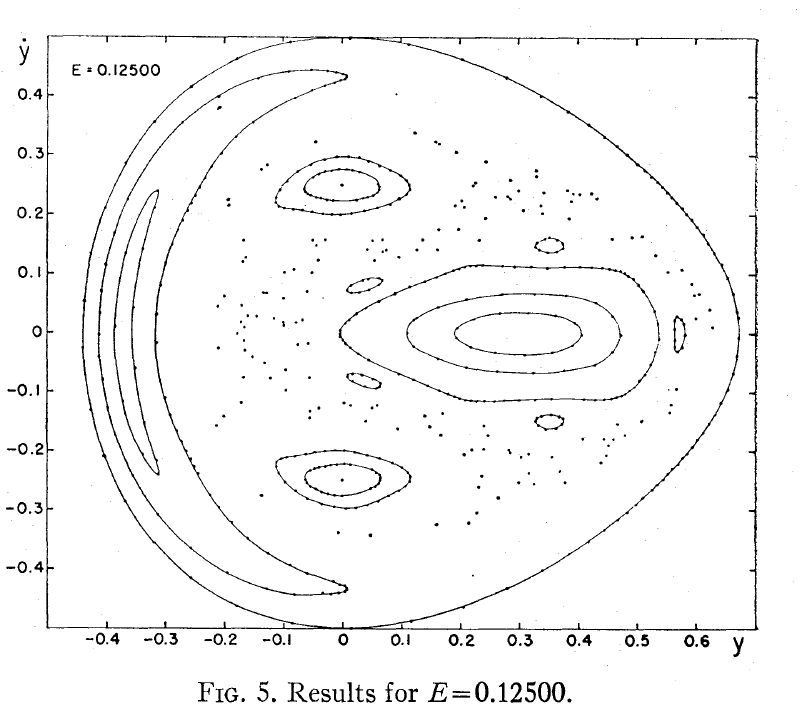} 
\caption{First numerical evidence of the intrication of chaotic and regular solutions in an Hamiltonian system  \citep{HenoHeil1964a}.
The motion of a star in the galactic potential is reduced to a two degree of freedom  Hamiltonian system, 
with a 4-dimensional phase space $(x,y,\dot x, \dot y)$. The energy is fixed, which reduces the  dimension 
of the allowed space to 3, and the successive  intersections of the trajectories with  the plane of coordinate 
$x=0$ are plotted (for $\dot x > 0$). In this Poincaré surface of section \citep{Poin1899a}, the curves 
are the successions of points belonging to regular trajectories that remain on invariant tori, as predicted by 
KAM theory. The clouds of dots are the successive intersections of chaotic trajectories with the 
surface of section.  
}
\label{fig1} 
\end{figure}

These numerical simulations were of great interest for Arnold who was at the time in Paris. It  was actually 
unusual for a mathematician to be interested by numerical simulations. For most of the 
mathematicians of the time in France, largely influenced by the   Bourbaki  rewriting 
of all mathematics in a very formal way,  numerical simulations were of little interest.
Arnold wrote a long letter to Hénon  (in french) that has been preserved and where he 
proposes some interesting questions related to the  numerical experiments 
of Michel Hénon (Fig. 2).  In these six pages he discusses various problems that could be 
addressed  by numerical methods. 

\begin{figure}[h!]
\includegraphics[width=12cm]{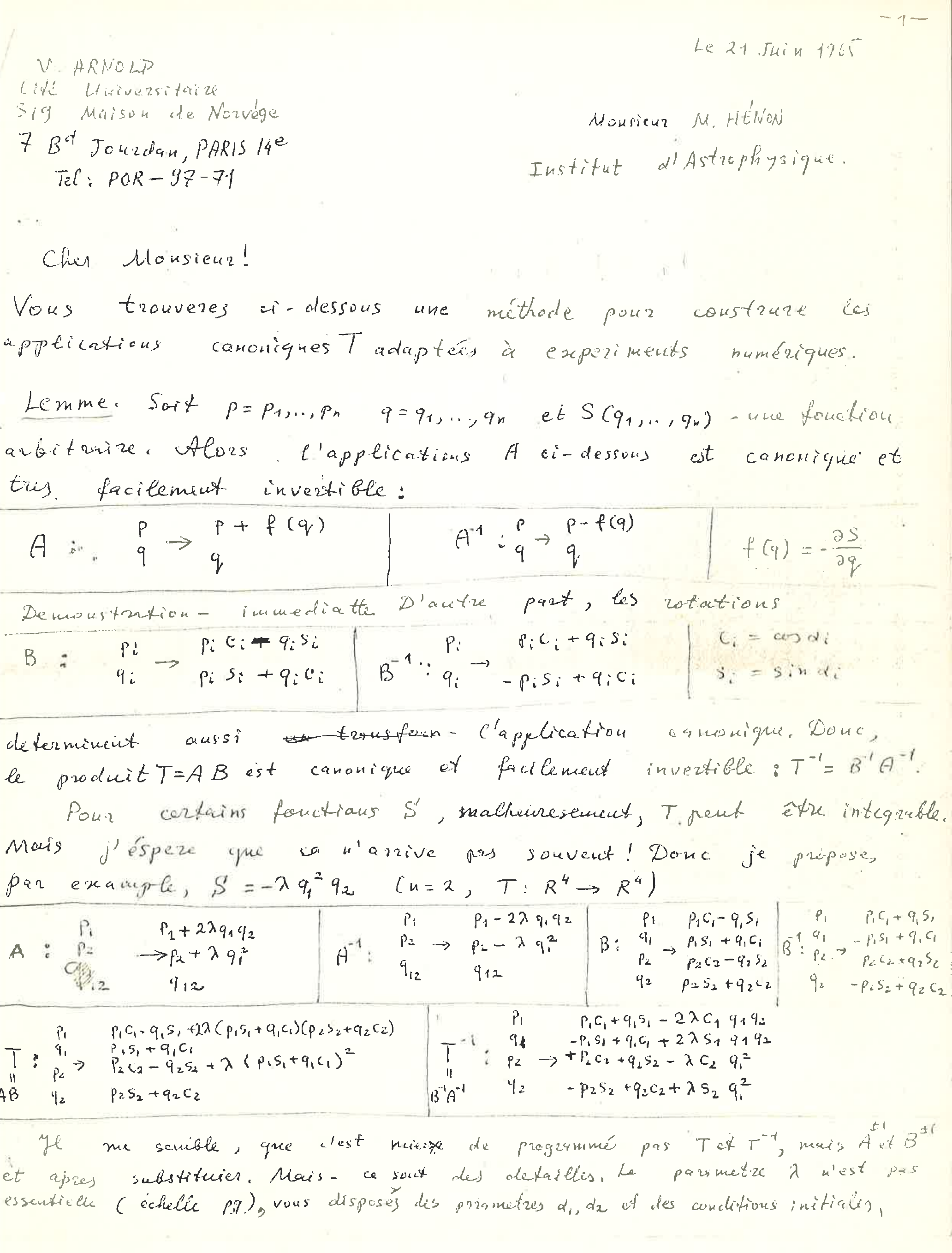} 
\caption{Letter of V.I. Arnold to M. Hénon  dated June, 21, 1965. 
Dear Sir ! You will find below a method for the construction of canonical applications, 
well adapted to numerical experiments \dots   
}
\label{fig2} 
\end{figure}

\section{Hénon's comment on Arnold's theorem}
In the archives of  M. Hénon, at the end of this long letter from Arnold, was attached 
a short comment  on Arnold's theorem on the planetary  problem (Fig.3).
This note can be transcribed  as follows

\medskip
\it
We have (ARNOLD, 1963, Usp. mat. Nauk., n$^\circ 5$, p.16)
\be
\EQM{
\delta^{(3)} &\leq   e^{2n} (32 n^2 + 100 n )^{-2n} \ ;  \crm
\delta^{(5)} &\leq   \delta^{(3)}
}
\ee 
and, p.23 
\be
\EQM{
\delta_1 <  \delta^{(5)} \ ;  \crm
M = \delta_1^{8n+24}  \ . \crm
}
\ee 
from where 
\be
M < \left( \Frac{e}{32n^2 + 100n} \right) ^{2n(8n+24)}
\ee
$M$ is thus \underline{extremely small} : for $n=2$, the smallest 
value of interest, we have :
\be
M < \left( \Frac{e}{328} \right)^{160} < 10^{-320}  \qquad  !!
\ee

Thus the theorem has only a theoretical interest and is absolutely not of practical use, 
at least in the presented form (cf p.18, §1.5, 23) : 
The perturbation needs indeed to be extremely small. 

As an example, the stability of the Solar System \underline{is not demonstrated}. 
\medskip

\rm

\begin{figure}[ht]
\includegraphics[width=12cm]{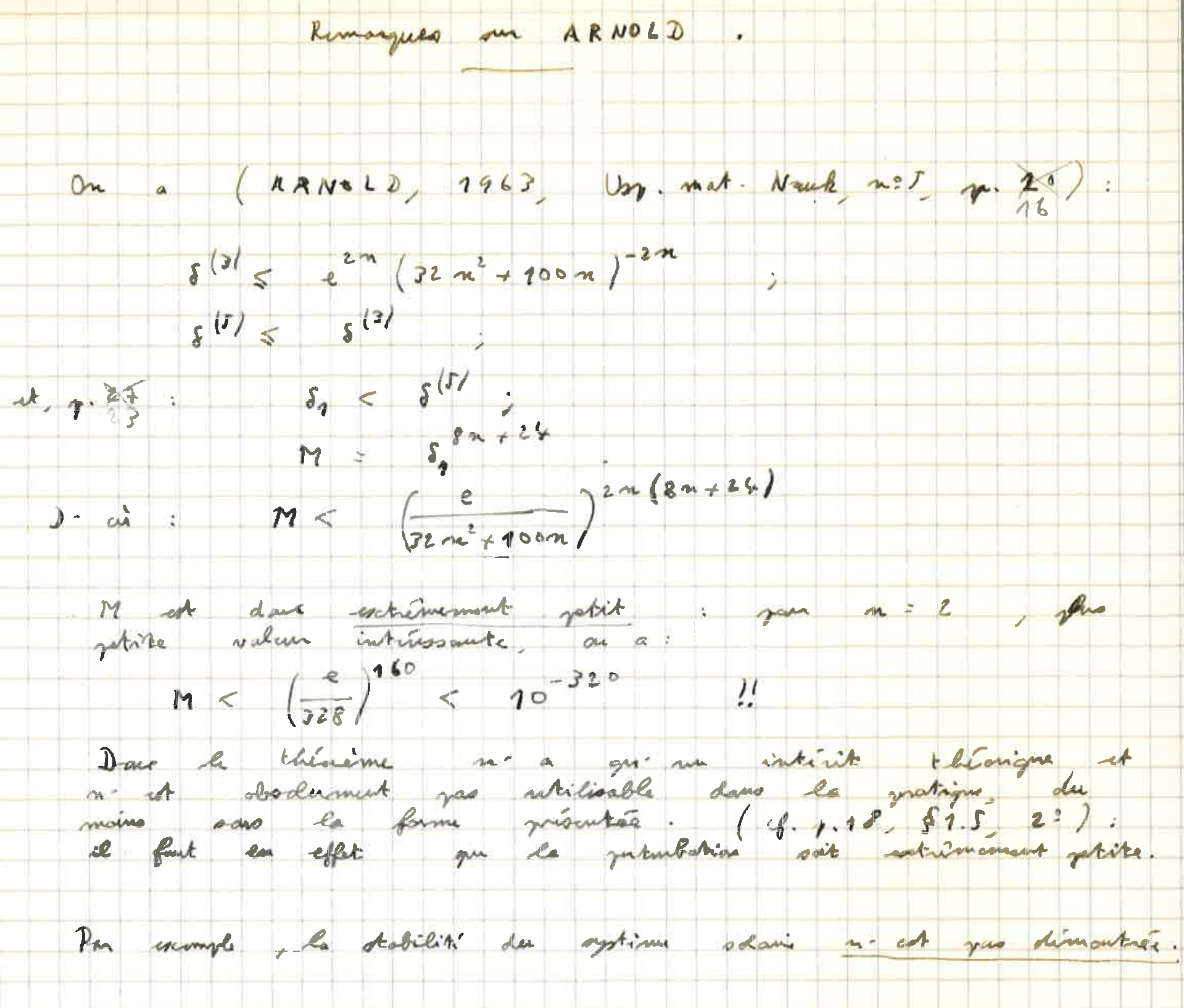} 
\caption{Note of Hénon,  joined to Arnold letter (Fig2) in Hénon's archives.}   
\label{fig3} 
\end{figure}

In this note, Hénon makes an estimate of  a bound on the maximal size of the perturbation 
that is allowed for the application of Arnold theorem. The value of this  bound, 
which can be thought as the ratio of the mass of the planet with respect to the Sun,  is extremely small. 
For  the most favourable case, with only two degrees of freedom, 
it provides a  constant of $10^{-320}$ ! 

When I started to study the long time behaviour of the Solar System, in the 80s, I was not aware 
of the  manuscript note of Michel Hénon, but I knew the result, as it  had been published 
in \citep{Heno1966a}, without the remark on the Solar System stability. 
He just stated there\footnote{
 Thus, these theorems, although of great theoretical interest, cannot, in their present condition,
  seem to be possibly applied to practical problems, where the perturbations are 
 always much larger \dots}
 
 {\it Ainsi, ces théorèmes, bien que d'un très grand intérêt théorique, ne semblent pas pouvoir 
 en leur état actuel être appliqués à des problèmes pratiques, où les perturbations sont 
 toujours beaucoup plus grandes \dots}

These statements were important for me as I read them in the late 80s, as they showed that 
the question of the stability of the Solar System was not solved. At the time, 
the main idea was that the Solar System  {\it is probably stable (by
any reasonable definition of the term) over time scales comparable with its age}
\citep{Murr1988a}. After the  results of Laplace and Lagrange on the stability 
at first order, many other "proofs" had come, the latest one being the demonstration of 
Arnold. At the time, the mathematicians were continuing their efforts, 
in order to provide rigorous proofs of this stability for a realistic model. But 
the estimate of Hénon showed that the path to these rigorous proofs was still long. 

\section{Arnold's paper}

Although I had read the comment of Michel Hénon, I had no clue on how he was able to derive his estimate 
from \citet{Arno1963a} paper. Indeed, this paper is very hard to read, 
and even  after  its study made during the thesis work of Philippe Robutel on the prolongation 
of Arnold's work \citep{Lask1995a,Robu1995a}, the derivation of Hénon was not clear. 
At the occasion of the present celebration of Michel Hénon, I went back to the paper of Arnold, 
and specifically searched for the estimate of Hénon. 
It appears that the derivation is actually relatively simple, once  the  derivation of 
Arnold is taken for granted. 
The bound to search for is the $M$ bound in equation (2) of Fig.4. In fact, following the notes of Hénon, 
it becomes easy to understand the derivation of the estimate. 
In page 16, we have 

\begin{figure}[ht]
\includegraphics[width=12cm]{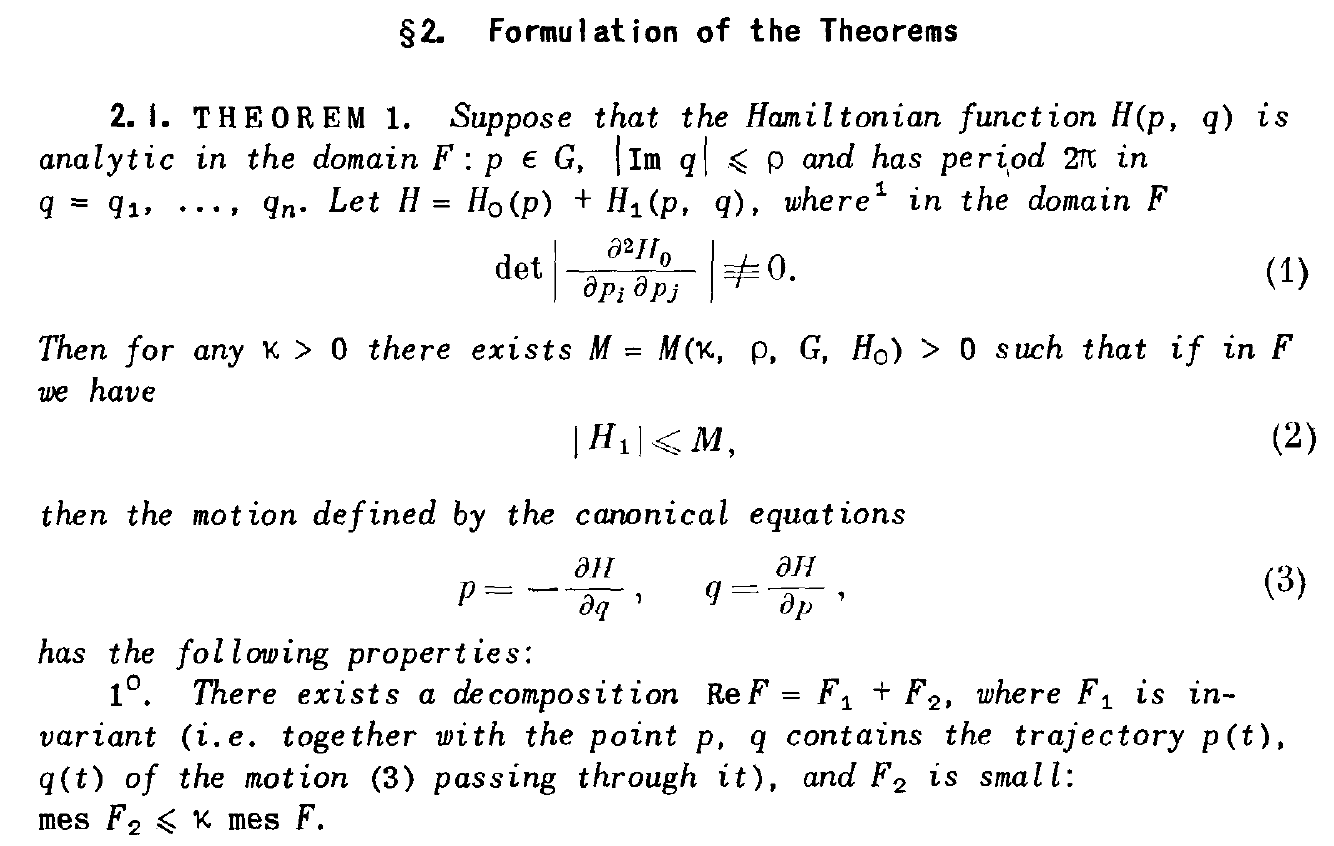}
\includegraphics[width=12cm]{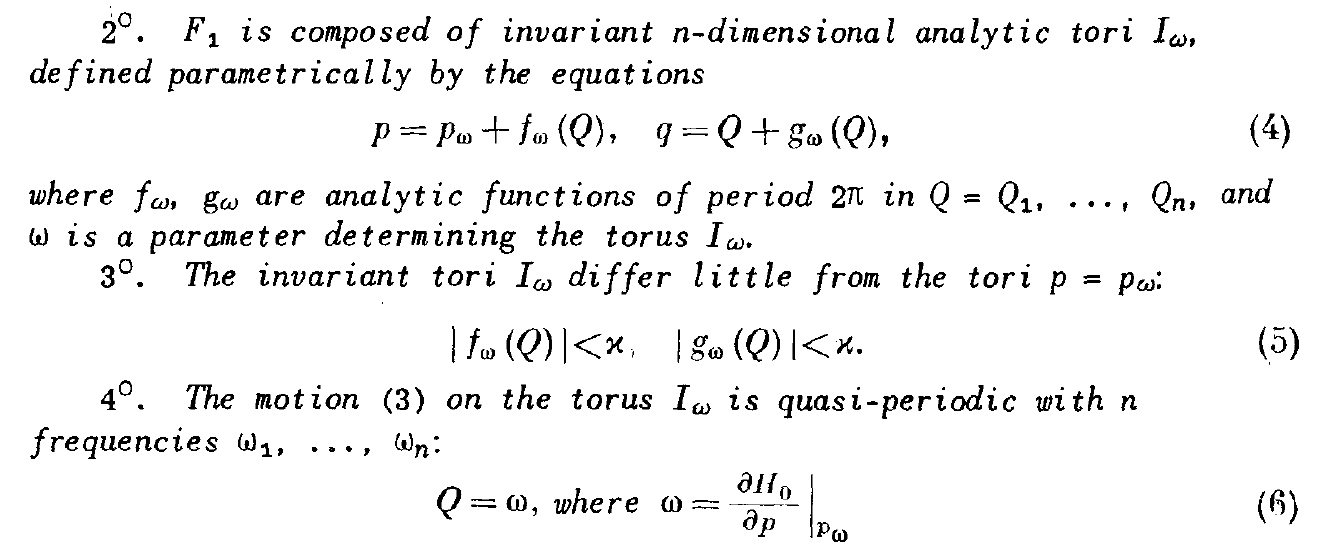} 
\caption{Arnold's theorem \citep{Arno1963a}.}   
\label{fig4} 
\end{figure}

\medskip
\noindent
\fbox{
\includegraphics[width=12cm]{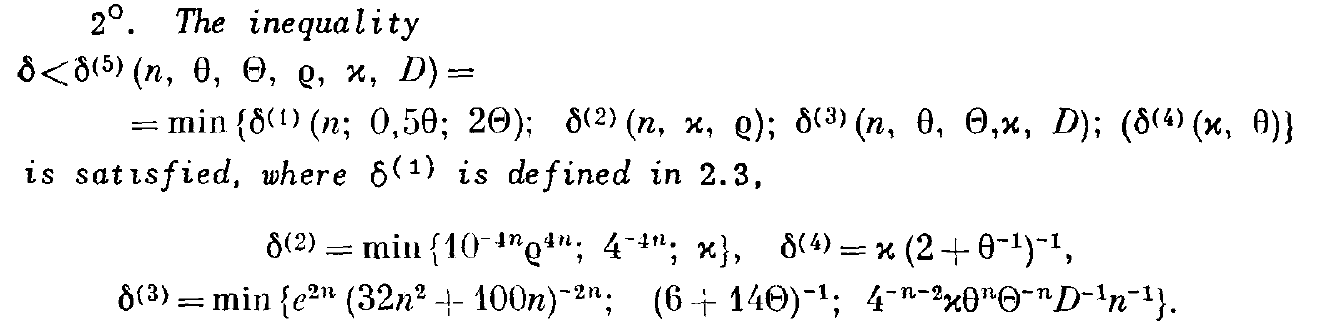} 
}
\medskip

Most of the  estimates of $\delta^{(3)}$ are complicated, but it is sufficient to consider the simplest 
one, which expresses easily in term of $n$, that is 
\be
\delta^{(5)} \leq \delta^{(3)} \leq e^{2n} (32 n^2 +100 n)^{-2n} \ .
\ee

Following Hénon, we have then  in page 23

\medskip
\noindent
\fbox{
\includegraphics[width=12cm]{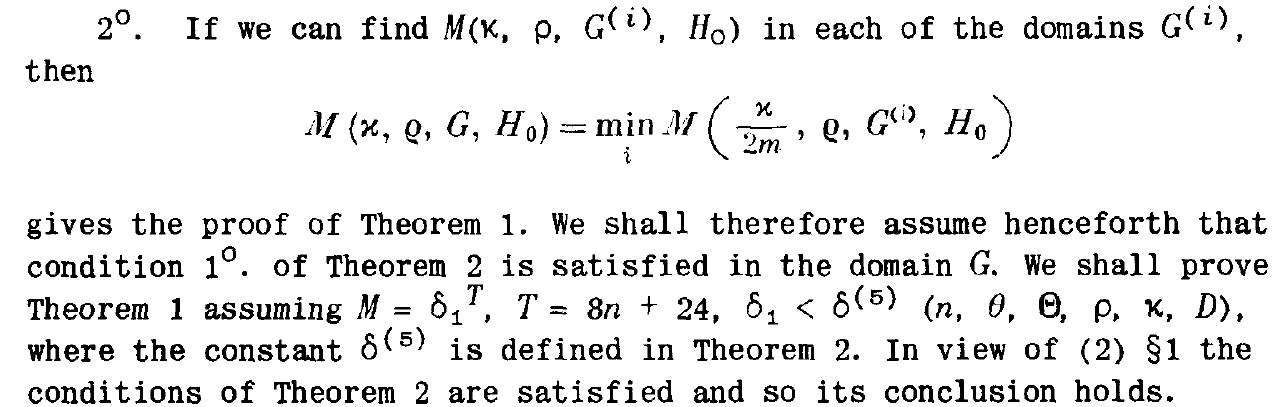}  
}

\medskip 
\noindent 
which gives $ M < {\delta^{(5)}}^{8n+24}$, that is 
\be
 M < \left( \Frac{e}{32n^2 + 100n} \right) ^{2n(8n+24)} \ ,
\ee
as reported by Hénon. The application for $n=2$ gives 
\be
M < 10^{-333.05}  
\ee
which is close to  the value $10^{-333}$ reported in \citep{Heno1966a}. This result was 
important, as it clearly showed that there was  a gap between the 
rigorously demonstrated results of stability  of the mathematicians and the 
real  Solar System. Indeed, if one only takes 3 degrees of freedom, the estimate becomes 
$10^{-672.}$. For a spatial planetary system limited to Jupiter and Saturn, 
after the reduction to heliocentric coordinates, and reduction  of the 
angular momentum,  there remain   four degrees of freedom \citep[e.g.][]{Robu1995a}
 ($n=4$), and thus  $M<10^{-1131.}$ .

\section{Chaos in the Solar System}
In  1981 was held in Les Houches a summer school devoted to the "Chaotic behaviour of deterministic 
systems".   I was not present to this conference, but I understand from 
the comments of many that this conference was an important landmark in the 
rising field of application of dynamical system theory to realistic physical systems. 
Although not present, I benefited from the chapter written by Michel on the 
numerical exploration of Hamiltonian systems \citep{Heno1983a} which I studied 
thoroughly as everything was there in order to understand the numerical experiments 
on chaotic systems. Present at {\it Les Houches} were in particular Alain Chenciner, 
 and also Jack Wisdom, who later on made a postdoctoral stay in Nice Observatory where 
 he   worked with Michel Hénon on the billiard dynamics \citep{Heno1983b}.   
 Later on, he  did the first study of the chaotic behavior of 
 the rotational motion of Hyperion, which was the first example of observable 
 chaotic behaviour in the Solar System \citep{Wisd1984a}. 
 
 As stated in the beginning of this text,  the better understanding of  the dynamical behaviour of 
 realistic models of physical systems was of fundamental importance 
 for understanding  that the motion of the planets themselves is chaotic 
 \citep{Lask1989a,Lask1990a}. Using dedicated computer algebra, I had been able 
 to average the equations of motion of the planets over their rapid motion, 
 and thus to construct,   following the works of Lagrange and Laplace, 
 a "secular" system of equations that represented the slow precessing motion 
 of the planetary orbits. The final system contained 153824 polynomial terms, but 
 as only the slow precessing motions were present, it could be integrated 
 very efficiently  with a very large step size of 500 years compared to 
 half of a day that was required by  a conventional integration of the equations of 
 motion. For several years, I had on hand the numerical output of the 
 integrated solutions over a few millions years, 
 but had difficulties to understand their complicated behaviour, 
  until I realised that 
 the system was actually chaotic, with an intricate   network of secular resonances 
 \citep{Lask1987a,Lask1989a,Lask1990a}.
 
In the 80s, there was still a huge gap between the 
small fraction of pioneer scientists like Michel Hénon and the majority 
of researchers in astronomy who were still thinking as if everything 
were regular, as if Henri Poincaré had never existed. The following 
personal recollection can be thought as an illustration of this statement.  
When I discovered that the Solar System is chaotic, I was still 
a young scientist, at the lower level of the CNRS carrier, with 
the title of "Chargé de recherche" of second class. After being for four 
years in this position, I could   apply for being upgraded to 
the  first class on the same position, which had to be decided by a national committee 
of astronomy experts. 
I met my referee who told me that there should not be any problem as my 
records were very good and  there was nine positions for only eleven applicants. 
Then, I told him that I had a new results, and that I had just shown that 
the Solar System is chaotic, and gave him the proofs of the Nature paper 
on this result, which was soon to be published \citep{Lask1989a}. The committee 
session occurred. Their global reaction was essentially to say that if it were true, 
it would be known, and that it is well-known  since Laplace 
that the Solar System is stable. I did not get the promotion
\footnote{Twenty five years later, I had the opportunity to discuss 
with a member of this committee, who acknowledged that 
they had not understood at all the meaning of the result, which 
 actually discredited me in their evaluation.}. 
Luckily, there were 
people like Michel Hénon,  and his former student Claude Froeschlé, in Nice, 
that clearly stated the importance and the meaning of the result, so the next year it was known\dots, 
and I got  promoted.  It should be said that meanwhile I had presented these results in 
front of some of the most prestigious mathematicians, as Jurgen Moser  who contributed 
to the KAM theory \citep{Mose1962a}. For them, it was not such a surprise, as 
Poincaré work and  KAM theory already provided the general setup of the 
story. I was just telling them that, instead of being very close to the 
central stable  quasiperiodic trajectory, which would be the cases for extremely small 
values of the planetary masses,  the actual Solar System was much further, in an area 
where resonances of large amplitude destroy a large amount of the regular trajectories,
as forecasted by Poincaré. 
Maybe it is at this meeting in Luminy in 1990, or in an earlier meeting
on dynamical systems, that Jurgen Moser 
spoke  in very high terms  of Michel to Alain Chenciner, saying {\it Michel Hénon is a real scientist}.

\printbibliography

\end{document}